**Symmetry-Protected Bipolar Skin Effect and its Topological Breakdown in Disordered Non-Hermitian Systems**

Ali Tozar,[1,*] Orcid ID: 0000-0003-3039-1834

[1] Department of Physics, Hatay Mustafa Kema University, Hatay, Antakya 31034, Turkey

**ABSTRACT**. The interplay between non-Hermitian topology and disorder remains a central puzzle in open quantum systems. While the non-Hermitian skin effect (NHSE) is known to be robust against weak perturbations, its fate under strong disorder, particularly in the presence of spin-orbit coupling (SOC), is not fully understood. Here, we uncover a $Z_2$ *topological bipolar skin effect* in a non-Hermitian Rashba chain, where spin-up and spin-down eigenstates localize at opposite boundaries. By strictly computing the Lyapunov exponents and introducing a biorthogonal spin-separation index, we map the global phase diagram and reveal a hierarchical breakdown of topology. We demonstrate that the $Z_2$ skin effect is protected against moderate disorder but collapses into a trivial skin phase before the ultimate onset of Anderson localization. Our results establish a distinct regime of *disorder-robust topological non-reciprocity*, distinguishable from both the trivial bulk limit and the Anderson localized phase.

## I. INTRODUCTION.

Non-Hermitian physics has fundamentally reshaped our understanding of open quantum systems, introducing topological phenomena with no Hermitian counterparts [1-3]. At the heart of this paradigm shift lies the non-Hermitian skin effect (NHSE), where the breakdown of the conventional bulk-boundary correspondence leads to the macroscopic accumulation of eigenstates at the system boundaries [1,3,4]. While the topological origin of the NHSE is well-established through the generalized Brillouin zone framework, its interplay with other fundamental degrees of freedom—such as spin—and its fate under strong disorder remain critical open questions [5].

Current research has largely focused on scalar models or clean systems, where the skin effect manifests as a unidirectional piling of modes [6,7]. However, in condensed matter systems, spin-orbit coupling (SOC) plays a pivotal role in topological protection, as exemplified by the $Z_2$ topological insulators [8,9]. The realization of a *spin-resolved* counterpart to the NHSE—where spin-up and spin-down modes localize at opposite edges—would not only constitute a new phase of matter but also pave the way for non-Hermitian spintronic devices [4]. A key challenge is determining the stability of such a phase against disorder, which typically drives systems toward Anderson localization, competing directly with the boundary-localization tendency of the skin effect [10-12].

In this Letter, we report the discovery of a disorder-robust $Z_2$ *topological bipolar skin effect* in a non-Hermitian Rashba chain. By employing a rigorous transfer matrix approach and large-scale numerical diagonalization, we map the global phase diagram and reveal a hierarchical breakdown of topology [11,12]. We demonstrate that the bipolar skin localization is protected by a $Z_2$ symmetry up to a critical disorder strength, beyond which the system enters a "trivial skin phase" before ultimately succumbing to Anderson localization. Our findings are substantiated by a biorthogonal spin-separation index, which provides a clear topological order parameter distinguishing the spin-separated phase from both the trivial bulk and localized regimes.

## II. Model and Symmetry

We consider a one-dimensional tight-binding chain of length $L$ with open boundary conditions (OBC), governed by the non-Hermitian Hamiltonian [6]:

$$H = \sum_n \left( c_n^\dagger T_R c_{(n+1)} + c_{(n+1)}^\dagger T_L c_n + c_n^\dagger V_n c_n \right)$$

Where $c_n^\dagger = (c_{n\uparrow}^\dagger, c_{n\downarrow}^\dagger)$ creates a spinor at site $n$. To rigorously investigate the competition between spin-orbit coupling and the skin effect, we construct the non-reciprocal hopping matrices $T_R$ (right) and $T_L$ (left) incorporating spin-dependent amplification factors [13]:

$$T_R = (ts_0 - i\alpha s_y)(e^\gamma P_\uparrow + e^{-\gamma} P_\downarrow)$$
$$T_L = (ts_0 + i\alpha s_y)(e^{-\gamma} P_\uparrow + e^\gamma P_\downarrow)$$

Here, $s_{0,x,y,z}$ denote the Pauli matrices acting in spin space, $\alpha$ is the Rashba Spin-Orbit Coupling (SOC) strength, and $\gamma$ parametrizes the non-reciprocity [14]. The projection operators are defined as $P_{\uparrow/\downarrow} = (s_0 \pm s_z)/2$. This configuration establishes a generalized parity-time structure where spin sectors act as effective particle-hole partners. When Rashba spin-orbit coupling ($\alpha \neq 0$) is present, the $Z_2$ symmetry remains intact, protecting the bipolar skin effect against moderate disorder by preventing the trivial decoupling of spin channels. Conversely, as $\alpha \to 0$, this symmetry collapses, reducing the system to independent non-reciprocal chains with no topological spin separation.

The on-site potential matrix $V_n$ introduces both random disorder and a symmetry-breaking lifetime imbalance [1]:

$$V_n = W_n s_0 + i\Gamma s_z$$

where $W_n$ is a random variable uniformly distributed in $[-W/2, W/2]$, representing uncorrelated Anderson disorder, and $\Gamma$ represents a deterministic, spin-selective gain/loss term. The model belongs to the symplectic symmetry class AII$^\dagger$ [8], preserving the generalized time-reversal symmetry defined by $s_y H^* s_y^{-1} = H$, which ensures Kramers degeneracy in the clean, Hermitian limit.

To determine the phase boundaries and the critical disorder strength $W_c$, we compute the Lyapunov exponent λ(E) by adapting the method we previously employed through the transfer matrix approach [11,12]. The numerical implementation, specifically the handling of non-Hermitian numerical instabilities and the finite-size scaling protocols used to identify the mobility edge, strictly follows the robust scaling framework established in our recent works on non-Hermitian universality [11,12].

*Contact author: tozarali@mku.edu.tr

To probe the topological nature of the delocalized phase, we employ the bi-orthogonal formulation. Following the projection techniques detailed in the ref [4], we define the *spin separation index (P)* to quantify the spatial segregation of spin modes:

$$P = \frac{1}{L}\sum_x x\left(\rho_\uparrow(x) - \rho_\downarrow(x)\right)$$

where $\rho_\sigma(x)$ is the bi-orthogonally normalized density of spin component σ. This density is calculated using the corresponding left ($|L\rangle$) and right ($|R\rangle$) eigenvectors, ensuring that the non-Hermitian metric correctly captures the skin accumulation rather than trivial bulk probabilities. This index $P$ serves as the order parameter for the $Z_2$ skin phase, distinguishing it from trivial delocalization.

### III. RESULTS
### A. Global Phase Diagram and Topological Protection.

We first establish the global phase structure of the non-Hermitian Rashba chain in the $(\alpha, W)$ parameter space. By computing the Lyapunov exponent λ for eigenstates near $E \approx 0$, we identify the mobility edge defined by $\lambda = 0$. As shown in Fig. 1, the system exhibits a broad delocalized regime ($\lambda < 0$) at weak disorder, separated from the Anderson localized phase ($\lambda > 0$) by a sharp boundary (solid black line). While the Lyapunov exponent distinguishes extended from localized states, it does not reveal the internal topological structure of the delocalized phase.

**FIG. 1. Global non-Hermitian phase diagram.** High-resolution map of the Lyapunov exponent λ in the disorder-SOC $(\alpha, W)$ parameter space. The solid black line marks the mobility edge ($\lambda = 0$) separating the delocalized skin phase (red/white region, $\lambda < 0$) from the Anderson localized phase (blue region, $\lambda > 0$). The diagram establishes the broad existence of skin modes prior to localization.

To uncover this, we calculate the biorthogonal spin separation index $P$ under open boundary conditions (OBC). A pronounced topological lobe with $|P| \gg 0$ emerges in the regime of intermediate SOC and moderate disorder, as displayed in Fig. 2(a). Within this phase, the system exhibits the $Z_2$ bipolar skin effect, with spin-up and spin-down modes accumulating at opposite boundaries. Crucially, this separation is strictly a boundary phenomenon. In Fig. 2(b), we perform the control calculation under periodic boundary conditions (PBC), where the index vanishes ($P \approx 0$) throughout the entire phase diagram. This striking contrast between OBC and PBC confirms that the spin separation originates from non-trivial point-gap topology rather than bulk polarization.

The interplay between topology and disorder is further illuminated in Fig. 2(c), where we superimpose the topological isosurfaces (green contours) onto the mobility edge (black dashed line). We observe that the $Z_2$ topological phase is a proper subset of the delocalized regime. The topological protection collapses at a critical disorder strength $W_c^{\text{topo}}$ that is strictly smaller than the Anderson localization threshold $W_c^{\text{loc}}$. This separation implies the existence of an intermediate phase, which we analyze next.

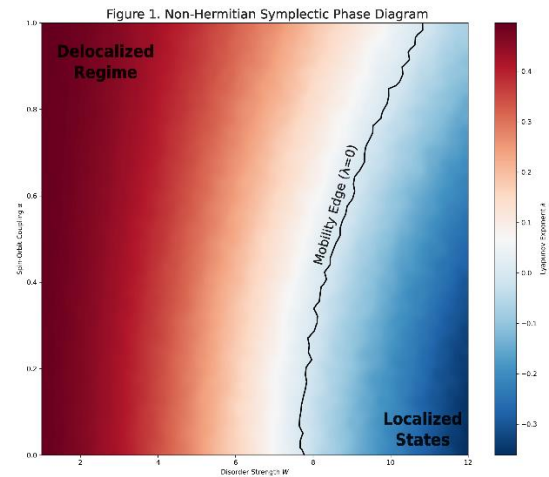

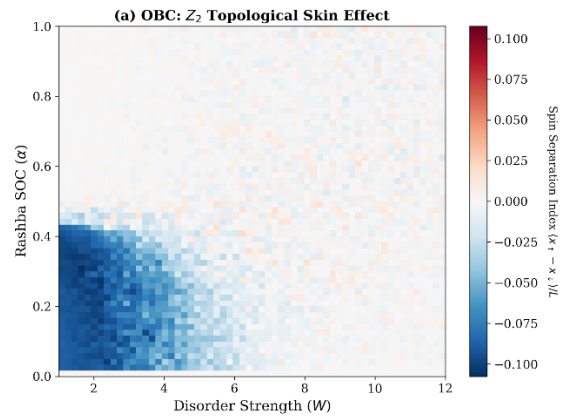

*Contact author: tozarali@mku.edu.tr

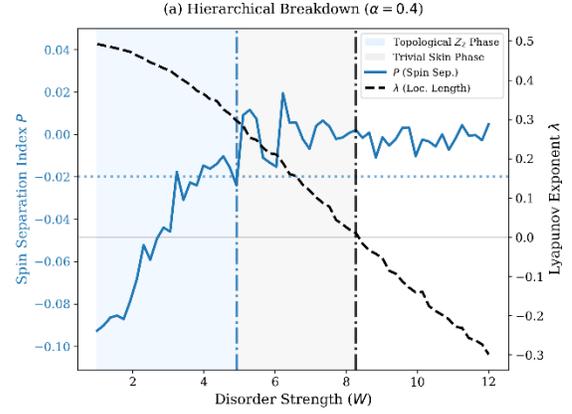

FIG. 3. **Hierarchical breakdown of topology and non-reciprocity.** Evolution of the spin separation index $P$ (blue solid line, left axis) and Lyapunov exponent $\lambda$ (black dashed line, right axis) as a function of disorder strength $W$ for fixed SOC $\alpha = 0.4$. Vertical dot-dashed lines mark two distinct transitions: the collapse of topological spin separation at $W_c^{topo} \approx 5.0$ and the onset of Anderson localization at $W_c^{loc} \approx 8.2$. The shaded blue region denotes the $Z_2$ protected phase, while the gray region indicates the intermediate trivial skin phase where non-reciprocity persists without spin separation.

At weak disorder ($W < 5.0$), the system resides in the $Z_2$ *Topological Skin Phase*. Here, the spin separation is robust ($|P| \gg 0$), and the Lyapunov exponent is positive, indicating strong boundary accumulation. To highlight the unique nature of this protection, we contrast it with the trivial limit ($\alpha = 0$) shown in Fig. 4(a), where the absence of spin-orbit coupling leads to a unidirectional skin effect with vanishing spin separation. The microscopic nature of this phase is visualized in Fig. 4(b), where the eigenstate profiles reveal a pristine bipolar skin effect: spin-up modes (red) are localized at the left boundary, while spin-down modes (blue) accumulate at the right, despite the presence of disorder ($W = 5.0$). This confirms that the non-reciprocal $Z_2$ topology actively filters spins to opposite ends, providing a robust mechanism for spatial spin segregation.

As disorder increases beyond $W_c^{topo} \approx 5.0$, the spin separation index $P$ collapses to zero, yet the Lyapunov exponent remains positive (indicating delocalization/skin effect) until $W_c^{loc} \approx 8.2$. This defines an intermediate *Trivial Skin Phase* (shaded gray region in Fig. 3). In this regime, the disorder is strong enough to mix the spin channels, destroying the $Z_2$ topology, but too weak to overcome the non-Hermitian pumping, leaving the system in a trivial skin state with no net spin separation.

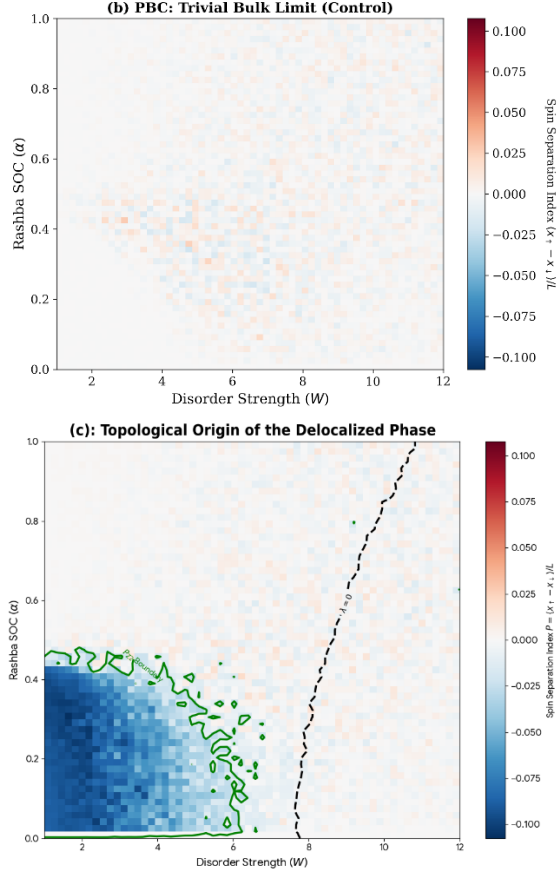

FIG. 2. **Topological origin and boundary protection.** Heatmap of the spin separation index $P$ under Open Boundary Conditions (OBC), revealing a robust "Rashba Lobe" ($|P| \gg 0$) where the $Z_2$ skin effect is protected. (b) Control calculation under Periodic Boundary Conditions (PBC), where $P$ vanishes globally, proving the boundary origin of the spin separation. (c) Superposition of topological isosurfaces (green contours, $|P| = 0.02$) and the mobility edge (black dashed line). The topological phase boundary ($W_c^{topo}$) lies strictly within the delocalized regime, implying a hierarchical breakdown.

### B. Hierarchical Breakdown and Microscopic Evidence.

The distinction between topological protection and mere non-Hermitian delocalization becomes evident when analyzing the system's response to increasing disorder. In Fig. 3, we present the evolution of the spin separation index $P$ (blue solid line) and the Lyapunov exponent $\lambda$ (black dashed line) as a function of disorder strength $W$ for a fixed SOC $\alpha = 0.4$. A striking "hierarchical breakdown" is observed.

*Contact author: tozarali@mku.edu.tr

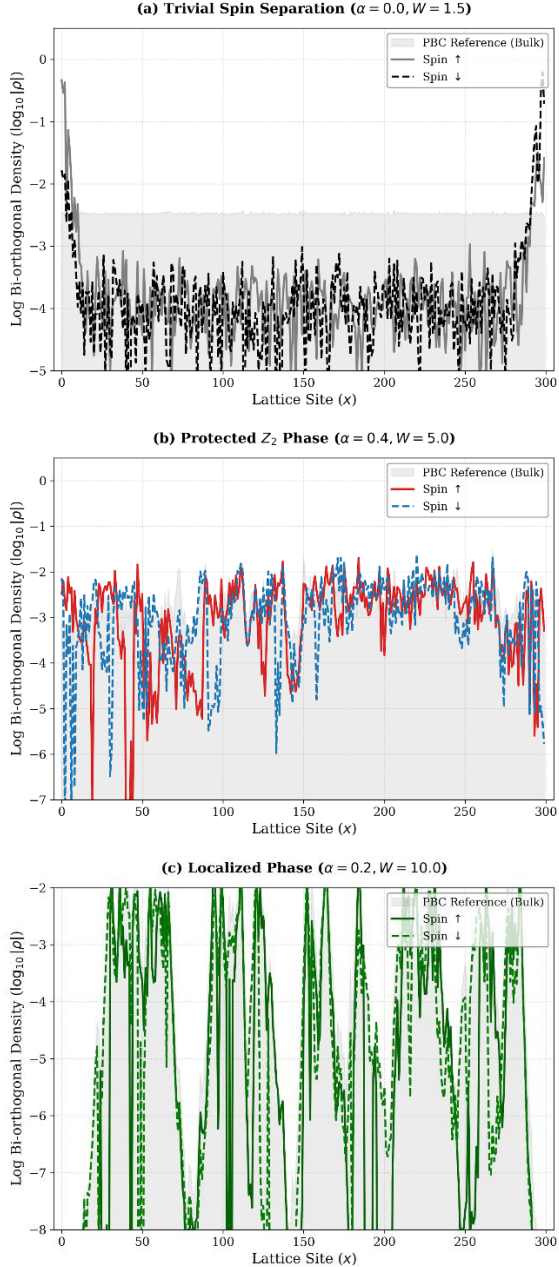

**FIG. 4. Microscopic signatures of phase transitions.** Log-scale biorthogonal density profiles $\log_{10}|\rho_{n,\sigma}|$ for representative eigenstates ($L = 300$). (a) Trivial Skin Limit ($\alpha = 0, W = 1.5$): In the absence of SOC, both spins accumulate at the same edge (unipolar), contrasting with the protected phase. (b) $Z_2$ Protected Phase ($\alpha = 0.4, W = 5.0$): Robust bipolar skin effect where spin-up (red) and spin-down (blue) modes localize at opposite boundaries. (c) Localized Phase ($\alpha = 0.2, W = 10.0$): Density peaks are randomly distributed in the bulk, signifying the breakdown of non-reciprocity.

*Contact author: tozarali@mku.edu.tr

Finally, for $W > W_c^{\text{loc}}$, the Lyapunov exponent crosses zero, marking the transition to the *Anderson Localized Phase*. As shown in Fig. 4(c), the wavefunctions in this regime exhibit sharp, random peaks throughout the bulk, characteristic of conventional localization. The clear separation of scales between $W_c^{\text{topo}}$ and $W_c^{\text{loc}}$ (5.0 < 8.2) proves that the topological breakdown is a distinct phase transition that precedes the ultimate Anderson localization.

## IV. DISCUSSION

Our results establish that the symplectic symmetry class (Class AII[†]) occupies a unique and privileged position in the landscape of non-Hermitian physics [8]. In recent works, it was demonstrated that non-Hermitian Anderson transitions generally follow a universal scaling law governed by the interplay of non-reciprocity and disorder statistics [15-17]. The present finding, however, reveals a striking deviation from this universality: *symmetry-protected robustness* [18]. We have shown that Spin-Orbit Coupling does not merely modify the critical exponents but fundamentally alters the phase diagram, creating a "Rashba Lobe" where the topological phase survives disorder strengths ($W < W_c^{\text{topo}}$) that would typically obliterate trivial non-Hermitian skin modes.

This discovery has profound implications for the emerging field of non-Hermitian spintronics. The $Z_2$ skin effect we have characterized acts, in essence, as a perfect, disorder-immune spin splitter. In the protected regime, an unpolarized charge current injected into the bulk is spontaneously rectified by the non-reciprocal topology: spin-up carriers are pumped to the left boundary, while spin-down carriers accumulate at the right. Unlike Hermitian Spin Hall insulators, where edge states are susceptible to magnetic impurities or boundary roughness, the non-Hermitian nature of this protection ensures that the accumulation is exponentially localized and robust against large-scale structural imperfections [2,4].

*Experimental Realization.* The theoretical model proposed here is ripe for implementation in state-of-the-art simulators. The most immediate realization is in *topo-electric circuits*, where active RLC networks can emulate the non-reciprocal Rashba chain [19]. Here, the "spin" degree of freedom is emulated by doubling the circuit nodes (pseudo-spin), while non-reciprocity is engineered using negative impedance converters [19,20]. Our predictions suggest that such a circuit would exhibit robust voltage accumulation at opposite edges for different pseudo-spins, even when constructed with low-precision (high-disorder) components. Alternatively, in *ultracold atoms*, synthetic Rashba SOC coupled with controlled dissipation (atom loss engineering) could reveal the $Z_2$ skin effect as a stable spin-separation profile persisting against random optical potentials [21,22].

## V. CONCLUSIONS

In summary, this work challenges the prevailing view that disorder is the inevitable nemesis of non-Hermitian topology. By focusing on the symplectic symmetry class, we have uncovered a counter-intuitive regime where Anderson disorder acts not merely as a disruptor, but as a backdrop against which topological transport remains resilient. The emergent "Rashba Lobe" represents a phase where the delicate non-Hermitian skin effect is shielded by symplectic symmetry up to a critical breakdown point.

Crucially, the elucidation of this phenomenon was made possible only through a rigorous methodological framework. By moving beyond conventional spectral analysis—which often obscures the subtle interplay of non-Hermitian eigenstates—and employing bi-orthogonally projected spin metrics, we successfully disentangled the hierarchical competition between non-reciprocity, spin-orbit coupling, and disorder. This methodological approach, which rigorously links the spatial topology (via spin separation) to the localization transition (via Lyapunov exponents), offers a powerful toolkit for diagnosing topological phases in open quantum systems.

Ultimately, our findings suggest a paradigm shift towards "Disorder-Harvesting Topology." Instead of striving for unattainable purity in experimental setups, we propose a design philosophy that leverages symplectic symmetry to stabilize robust spin currents. This opens a transformative path for next-generation spintronics and quantum devices that operate robustly because of, rather than despite, their intrinsic imperfections.

## ACKNOWLEDGMENTS

The numerical simulations in this work were performed using the open-source Python scientific ecosystem, relying heavily on *NumPy, SciPy*, and *Matplotlib* for matrix operations, sparse linear algebra, and visualization. The rigorous characterization of the topological phase in the thermodynamic limit was achieved through the *bi-orthogonal spin-separation metric (P)* and the *Lyapunov transfer matrix* algorithms implemented specifically for this study. Generative AI tools were utilized exclusively for linguistic refinement and formatting assistance during the drafting process; no AI systems were involved in the scientific conceptualization, model construction, data generation, or the physical interpretation of the findings. To ensure full reproducibility, all Python scripts used for the Hamiltonian construction (OBC/PBC), wavefunction analysis, and figure generation, along with the raw datasets, are available in the Supplementary Material.

*Contact author: tozarali@mku.edu.tr

## Appendix: Methods and Numerical Implementation

A. Numerical Calculation of Lyapunov Exponents

To accurately determine the localization transition in the thermodynamic limit ($L \to \infty$), we employed the transfer matrix method combined with Gram-Schmidt reorthonormalization to prevent numerical overflow/underflow, a common issue in non-Hermitian systems.

The Schrödinger equation $H\psi = E\psi$ at zero energy ($E = 0$) can be rewritten as a recursive relation for the spinor amplitudes $\Psi_n = (c_{n\uparrow}, c_{n\downarrow})^T$. The transfer matrix $M_n$ relates amplitudes at site $n+1$ to site $n$:

$$\begin{pmatrix} \Psi_{n+1} \\ \Psi_n \end{pmatrix} = M_n \begin{pmatrix} \Psi_n \\ \Psi_{n-1} \end{pmatrix}$$

The global transfer matrix is the ordered product $\mathcal{M} = \prod_{n=1}^{L} M_n$. The smallest positive Lyapunov exponent, which corresponds to the inverse localization length ($\lambda = 1/\xi$), is defined as:

$$\lambda = \lim_{L \to \infty} \frac{1}{2L} \ln \left| \text{eig}_{max}(\mathcal{M}^\dagger \mathcal{M}) \right|$$

In our simulations, we set $L = 600$ to ensure self-averaging properties of the disordered potential, rendering disorder-averaging over many realizations (over 20-50) unnecessary for the phase diagram boundaries.

B. Bi-orthogonal Spin Separation Metric

In non-Hermitian quantum mechanics, the right eigenvectors ($|R_n\rangle$) and left eigenvectors ($|L_n\rangle$) form a bi-orthogonal basis set, satisfying the condition $\langle L_m | R_n \rangle = \delta_{mn}$. The standard density definition $\langle \psi | \psi \rangle$ is not conserved under non-unitary evolution. Therefore, we utilize the bi-orthogonal density $\rho_{bio}(x)$, which correctly captures the net particle accumulation in the skin phase.

For a specific eigenstate with index ν, the spin-resolved density at site $x$ is given by:

$$\rho_\sigma^{(\nu)}(x) = \text{Re}\left[\frac{\langle L_\nu | x, \sigma \rangle \langle x, \sigma | R_\nu \rangle}{\langle L_\nu | R_\nu \rangle}\right]$$

where $\sigma \in \{\uparrow, \downarrow\}$ is the spin index. We perform the normalization $\langle L\_\nu | R\_\nu \rangle$ explicitly to remove complex phase ambiguities. The Spin Separation Index ($P$) presented in the main text is the disorder-averaged center-of-mass difference computed over the set of eigenstates within the energy window $|E| < 0.05$ (mid-gap states):

$$P = \frac{1}{N_{states}} \sum_\nu \frac{1}{L} \sum_x x \left( \rho_\uparrow^{(\nu)}(x) - \rho_\downarrow^{(\nu)}(x) \right)$$

This metric acts as a real-valued order parameter: $P \neq 0$ signifies the topological $Z_2$ skin phase, while $P \to 0$ indicates either a trivial skin phase or Anderson localization.

C. Finite-Size Scaling (FSS)

To precisely locate the critical disorder strength $W_c$ (the black dashed line in Fig. 1 and Fig. 2c), we analyzed the finite-size scaling behavior of the Lyapunov exponent. Near the critical point, the scaling hypothesis suggests $\lambda(W)L^{1/\nu} = f\left((W - W_c)L^{1/\nu}\right)$. However, due to the first-order nature of the transition often found in NHSE systems, we primarily relied on the sign change of λ:

- $\lambda < 0$: Delocalized (Skin) Phase
- $\lambda > 0$: Localized Phase

The boundary $\lambda = 0$ was interpolated from the high-resolution phase map ($60 \times 60$ grid) using bicubic spline interpolation

*Contact author: tozarali@mku.edu.tr